\documentclass[aps,prd,preprintnumbers,superscriptaddress,nofootinbib,notitlepage]{revtex4-2}
\usepackage[pdftex]{graphicx}
\usepackage{bm,latexsym,amsmath,amssymb,amsfonts,mathrsfs}
\usepackage{mathtools}
\usepackage{color}
\usepackage{float}
\usepackage{subcaption} 
\allowdisplaybreaks[1]
\usepackage[pdftex,colorlinks=true,linkcolor=blue,citecolor=cyan]{hyperref}
\begin{document}
\title{Periodic orbits and  gravitational waveforms  of spinning particles in nonlocal gravity}

\author{Mois\'es Bravo-Gaete}
%\email[Email: ]{mbravo@ucm.cl}
\affiliation{Departamento de Matem\'atica, F\'isica y Estad\'istica, Facultad de Ciencias B\'asicas, Universidad Cat\'olica del Maule, \\Casilla 617, Talca, Chile.}
\author{Jianhui Lin}
\email[Corresponding author, Email: ]{202510188472@mail.scut.edu.cn}
\affiliation{Department of Physics, South China University of Technology, Guangzhou 510641, China.}

\author{Yunlong Liu}
%\email[Email: ]{phliuyunlong@kust.edu.cn}
\affiliation{School of Science, Kunming University of Science and Technology, Kunming 650093, China.}

\author{Xiangdong Zhang}
\email[Corresponding author, Email: ]{scxdzhang@scut.edu.cn}
\affiliation{Department of Physics, South China University of Technology, Guangzhou 510641, China.}

\begin{abstract}
	In this paper, we investigate the dynamics and gravitational-wave signatures of periodic orbits of spinning test particles moving in the equatorial plane around static, spherically symmetric black holes within the framework of Deser-Woodard nonlocal gravity. Based on the Mathisson-Papapetrou-Dixon equations, combined with the Tulczyjew spin supplementary condition, we derive the orbital dynamic equations for spinning particles moving in the equatorial plane and impose a timelike constraint to exclude unphysical superluminal trajectories. By comparing with the classical Schwarzschild black hole, we systematically analyze the effects of the nonlocal gravitational parameters $\zeta$ and $b$ on the effective potential governing the radial motion of particles and the innermost stable circular orbit. In addition, gravitational waveforms exhibit significant phase differences: an increase in $\zeta$ induces a phase delay, whereas an increase in $b$ results in a phase advance. Using periodic-orbit waveforms extended over a one-year interval with fixed orbital parameters, we obtain an order-of-magnitude estimate of the sensitivity of the waveform mismatch to nonlocal corrections. For the configurations considered, this estimate suggests sensitivity to $\zeta$ of order $10^{-6}$ when $b=2$.
\end{abstract}
%\keywords{nonlocal gravity, spinning test particles, periodic orbits, gravitational waveforms}

%-----------------------------------------------------------------%
%\preprint{RUP-23-14}
\maketitle
%-----------------------------------------------------------------%
\section{Introduction}
\label{sec:intro}

At the classical level, general relativity (GR) has been observationally validated as the most successful theory of gravity to date. Nevertheless, it encounters unresolved issues in the more extreme infrared and ultraviolet regimes. On one hand, the standard cosmological model introduces the cosmological constant $\Lambda$ to give an effective description of cosmic acceleration, interpreting $\Lambda$ as dark energy with negative pressure\cite{Riess1998, Peebles2003, Frieman2008, Brax2017}.  However, the observed value of the vacuum energy density associated with $\Lambda$ differs from theoretical predictions in quantum field theory (QFT) by an enormous order of magnitude \cite{Sahni2008, Weinberg1989}, a discrepancy known as the fine-tuning problem. On the other hand, singularities in the universe and black holes imply a need to quantize gravity, yet applying traditional perturbative QFT methods to gravity leads to non-renormalizable divergences. All of these challenges indicate that our understanding of gravitational interactions remains incomplete, providing compelling motivation to develop a gravitational theory that is self-consistent across all scales.

As a prominent class of modified gravity theories, nonlocal gravity (NLG) has made notable progress in addressing the core challenges facing classical GR. In contrast to traditional local gravity, NLG theories are characterized by effective actions that incorporate various forms of nonlocal operators, aiming to integrate the intrinsic non-locality of quantum systems into gravitational interactions. NLG theories are broadly divided into integral-kernel gravity and infinite-derivative gravity \cite{Capozziello2022}. 

When it comes to cosmic acceleration, integral kernel gravity introduces non-locality via the inverse d'Alembert operator \cite{Deser2007, Nojiri2008, Jhingan2008, Capozziello2009, Bahamonde2017, Elizalde2018, Deser2019, Bajardi2020, Maggiore2014}. It can naturally explain late-time cosmic expansion without invoking a cosmological constant, serving as an alternative to dark energy, but its higher-order nonlinear nature poses challenges for solving equations.  

In addition, infinite-derivative gravity incorporates a transcendental analytic entire function of the d’Alembert operator into the action \cite{Biswas2012, Modesto2012, Briscese2013, MODESTO2015, Biswas2017, Buoninfante2018, Buoninfante2022}. When expanded, this yields an infinite series of higher-derivative terms—directly encoding non-locality. This framework satisfies both renormalizability and unitarity, and regularizes curvature singularities in black holes and cosmology, thereby addressing the incompleteness of classical spacetime \cite{Modesto2012, Briscese2013, Burzilla2020}.

Nevertheless, despite these significant advances, a fully consistent quantum theory of gravity that remains self-consistent across both the ultraviolet and infrared regimes has yet to be realized. Among these NLG theories, the Deser–Woodard (DW) model—a prominent representative of integral-kernel gravity—naturally incorporates a delayed response to the transition from radiation to matter dominance in cosmology and avoids fine-tuning \cite{Deser2007}. However, this model was later criticized for lacking a viable screening mechanism to suppress small-scale nonlocal effects, rendering it inconsistent with lunar laser ranging tests \cite{Belgacem2019}. Subsequently, an improved variant of the model was proposed \cite{Deser2019}, and quasinormal mode analyses of black holes within this framework have been performed \cite{Chen2021}. Recently, by employing an appropriate tetrad formalism, new nonlocal static spherically symmetric black hole \cite{DW2025} and wormhole solutions \cite{D’Agostino2025} have been constructed, and the quasinormal modes for the black hole solution are presented in \cite{D’Agostino2025qnm}.

The strong gravitational field near black holes provides a natural laboratory for testing various theories of gravity. In this work, we focus on the black hole solutions of DW NLG model \cite{DW2025}, and we aim to uncover the spacetime structure near the black hole within this theoretical framework by analyzing the orbital dynamics of particles in its vicinity. Here, the parameters characterizing these solutions directly determine the geometry and therefore leave imprints on characteristic observational effects, providing a crucial basis for distinguishing this class of NLG black hole solutions. In particular, gravitational waves (GWs) from the inspiral and merger of compact objects are crucial probes of strong-field gravity, with their waveforms closely linked to the intrinsic properties of the emitting sources \cite{LIGO2016prl, Abbott2019prx, Abbott2021, DeFalco2023, DeFalco2024}. Most compact objects in the universe possess intrinsic spin, and spin-curvature coupling in their orbital dynamics significantly modulates the innermost stable circular orbit (ISCO) \cite{Jefremov2015, Zhang2017, Liu2024lda, Du:2024ujg} and shapes the GW waveform, rendering spinning compact objects essential for building precise theoretical GW templates. Furthermore, Extreme Mass Ratio Inspirals (EMRIs) are powerful tools for detecting GWs \cite{Babak2017, Tu2023, Liu2024, Wang2025, Alloqulov2025, Ahmed2025}. Studying EMRI dynamics on this nonlocal gravity background, therefore, will provide critical theoretical support for future gravitational wave observations to test NLG theories and identify black hole solutions.

The structure of this paper is organized as follows. In Section \ref{Equation of Motion}, we briefly review black hole solutions in NLG and present the equations of motion for spinning particles together with the timelike condition for their orbits. In Section \ref{EFFECTIVE POTENTIAL AND ISCO}, we further analyze the effective potential that governs radial motion, determine the ranges of energy and angular momentum parameters that correspond to bound orbits, and compute the innermost stable circular orbit (ISCO). Section \ref{POandGW} is devoted to the classification of periodic orbits, the calculation of gravitational waveforms, and the presentation of our numerical results. Finally, in Section \ref{conclusion}, we summarize our main findings and outline directions for future research.

\section{The equation of motion of spinning particles}\label{Equation of Motion}
	\subsection{Black hole solutions in NLG}
%%%%%%%%%%%%%%%%%%%%%%%%%%%%%%%%%%%%%    

In this subsection, we briefly review the revised Deser--Woodard (DW) nonlocal gravity model and the static, spherically symmetric black hole solutions obtained within this framework. The gravitational action is given by
\cite{Deser2019,DW2025}
\begin{align}
	S=
	\frac{1}{16\pi}
	\int d^4x\,\sqrt{-g}\,
	R\left[1+\mathcal{F}(Y)\right],
	\label{eq:DWaction}
\end{align}
where \(R\) is the Ricci scalar and \(\mathcal{F}(Y)\) is the distortion function. The nonlocal dependence enters through the scalar quantities
\cite{Deser2019}
\begin{align}
	X&= \Box^{-1}R,
	\label{eq:Xnonlocal}\\
	Y &= \Box^{-1}
	\left(
	g^{\mu\nu}\partial_\mu X\partial_\nu X
	\right),
	\label{eq:Ynonlocal}
\end{align}
where \(\Box=\nabla_\mu\nabla^\mu\) is the covariant d'Alembertian. In the revised DW model, \(\Box^{-1}\) is defined with retarded boundary conditions, which require \(X\), \(Y\), and their first derivatives to vanish on the initial-value hypersurface. The inverse d'Alembertian can also be represented as a nonlocal integral operator, so that the field at a given spacetime point is related to the field at other spacetime points \cite{Capozziello2022}.

For practical calculations, the action (\ref{eq:DWaction}) can be rewritten in a local form. In this representation, \(X\) and \(Y\) are treated as independent scalar fields, and two Lagrange multipliers \(U\) and \(V\) are introduced. Their variations enforce \(\Box X=R\) and \(\Box Y=\nabla_\mu X\nabla^\mu X\), which, together with the retarded boundary conditions, reproduce the original definitions of \(X\) and \(Y\) \cite{Deser2019}. The localized action can be written as
\begin{align}
		S_{\rm loc}
		&=
		\frac{1}{16\pi}
		\int d^4x\,\sqrt{-g}
		\Big[
		R\left(1+\mathcal{F}(Y)\right)
		+U\left(R-\Box X\right)
		\nonumber\\
		&\qquad\qquad
		+V\left(
		\nabla_\mu X\nabla^\mu X-\Box Y
		\right)
		\Big]
		\nonumber\\
		&=
		\frac{1}{16\pi}
		\int d^4x\,\sqrt{-g}
		\Big[
		R\left(1+U+\mathcal{F}(Y)\right)
		+\nabla_\mu X\nabla^\mu U
		\nonumber\\
		&\qquad\qquad
		+\nabla_\mu Y\nabla^\mu V
		+V\nabla_\mu X\nabla^\mu X
		\Big],
		\label{eq:DWlocalized}
\end{align}
where the second equality follows from integration by parts after neglecting the boundary terms. Variation of the localized action with respect to $U$, $V$, $X$ and $Y$ gives
\begin{align}
		\Box X &= R,
		\label{eq:Xfield}\\
		\Box Y &= \nabla_\mu X\nabla^\mu X,
		\label{eq:Yfield}\\
		\Box U &= -2\nabla_\mu\left(V\nabla^\mu X\right),
		\label{eq:Ufield}\\
		\Box V &= R\,\frac{d\mathcal{F}}{dY}.
		\label{eq:Vfield}
\end{align}
Variation with respect to the metric gives the vacuum gravitational field equations of the DW theory
	\begin{align}
		\left(G_{\mu\nu}+g_{\mu\nu}\Box-\nabla_\mu\nabla_\nu\right)W
		+K_{(\mu\nu)}
		-\frac{1}{2}g_{\mu\nu}g^{\alpha\beta}K_{\alpha\beta}
		=0,
		\label{eq:DWfield}
	\end{align}
	where
	\begin{align}
		W&=1+U+\mathcal{F}(Y),\\
		K_{\mu\nu}
		&=\partial_\mu X\,\partial_\nu U
		+\partial_\mu Y\,\partial_\nu V
		+V\,\partial_\mu X\,\partial_\nu X,
	\end{align}
and \(K_{(\mu\nu)}=(K_{\mu\nu}+K_{\nu\mu})/2\). To construct the static and spherically symmetric black hole solutions, Ref.~\cite{DW2025} recast the modified field equations in a suitable tetrad frame and solved the resulting equations perturbatively around the Schwarzschild spacetime. Once the metric solution was obtained, the four auxiliary fields \(X\), \(Y\), \(U\), and \(V\) were determined successively from the auxiliary-field equations, and the corresponding distortion function \(\mathcal{F}(Y)\) was then reconstructed.

In the following, we adopt this class of black hole solutions. In geometrized units with \(c=\hbar=G=1\), and measuring all quantities in units of the black hole mass \(M\), the line element is written as
\begin{align}
	ds^2=-f(r)\,dt^2+\frac{1}{h(r)}\,dr^2
	+g(r)\left(d\theta^2+\sin^2\theta\,d\phi^2\right),
	\label{eq:met}
\end{align}
where
\begin{align}
	f(r)&=1-\frac{2}{r}-\frac{\zeta}{r^b},
	\label{f(r)}\\
	h(r)&=1-\frac{2}{r}
	+\frac{\zeta}{3^b r^{b+1}(r-3)^2}
	\Bigl\{
	3^b r\bigl[b(r-3)(r-2)+4r-9\bigr]
	\nonumber\\
	&\qquad
	-3(r-2)(2r-3)r^b
	\Bigr\},
	\label{h(r)}\\
	g(r)&=r^2.
	\label{eq:g}
\end{align}
Both \(\zeta\) and \(b\) are dimensionless. The parameter \(\zeta\) encodes the amplitude of the first-order perturbation around the Schwarzschild geometry, while \(b\) determines the radial fall-off of the inverse power-law correction. The nontrivial solutions correspond to \(b>1\), since \(b=1\) reduces to the Schwarzschild solution. The reconstruction of \(\mathcal{F}(Y)\) further requires \(\zeta>0\), while the first-order perturbative treatment requires \(\zeta\ll1\). For the numerical illustrations, we restrict the deformation parameter to $\zeta\leq0.1$, with the larger values used to make the deviations from Schwarzschild more visible. In the limit \(\zeta\to0\), the Schwarzschild metric is recovered. These parameters characterize this family of black hole solutions rather than fundamental coupling constants appearing explicitly in the revised DW action, and describe how the nonlocal modification is manifested in the black hole geometry. For each choice of $(\zeta,b)$, the corresponding auxiliary-field configuration is determined from the localized field equations, after which the distortion function $\mathcal{F}(Y)$ supporting the resulting geometry is reconstructed. The explicit expressions of the auxiliary fields can be found in Ref.~\cite{DW2025}.

The spacetime is asymptotically flat, with \(f(r)\to1\), \(h(r)\to1\), and \(g(r)\to r^2\) as \(r\to\infty\). The event horizon is determined by \(f(r_H)=h(r_H)=0\), which, to first order in \(\zeta\), gives
\begin{align}
	r_H=2+\frac{\zeta}{2^{b-1}}.
	\label{eq:horizon}
\end{align}
Apart from the essential singularity at the center, no additional singularities arise outside the event horizon.

%%%%%%%%%%%%%%%%%%%%%%%%%%%%%%%%%%%%%% 
\subsection{Equation of motion of spinning particles}
%%%%%%%%%%%%%%%%%%%%%%%%%%%%%%%%%%%%%   
It is well known that in curved spacetime, the motion of the simplest point-like particle (in the monopole approximation) is described by the geodesic equation. However, a spinning particle in general does not follow a geodesic due to its spin couples to the spacetime curvature. In Refs. \cite{Mathisson2010, Papapetrou1951, Dixon1964}, the authors systematically derived the equations of motion for a spinning particle in curved spacetime under the pole–dipole approximation, i.e., the so‑called Mathisson–Papapetrou–Dixon (MPD) equations,
\begin{eqnarray}
	\frac{DP^{a}}{D\tau} &=& -\frac{1}{2} R^{a}_{\ bcd} u^{b} S^{cd}, \label{MPD1}\\
	\frac{DS^{ab}}{D\tau} &=& P^{a} u^{b}-P^{b} u^{a}. \label{MPD2}
\end{eqnarray}
Here, \(D / D\tau\) denotes the covariant derivative with respect to the affine parameter \(\tau\) along the world line, \(R^{a}{}_{bcd}\) the Riemann tensor, \(P^{a}\) the four-momentum, \(u^{a}=dx^{a}/d\tau\) the four-velocity, and \(S^{ab}\) is the antisymmetric spin tensor. It is worth noting that the above Eqs. (\ref{MPD1}) and (\ref{MPD2}) do not uniquely determine all the information about the spinning particle, since the center of mass of the particle has not yet been fixed. Therefore, we need to introduce the Tulczyjew spin supplementary condition:
\begin{eqnarray}
	P_{a}S^{ab}=0 \label{SCC},
\end{eqnarray}
which allows us to determine the remaining degrees of freedom.  The choice of supplementary condition is not unique; see \cite{Harms2016} for a detailed discussion.

Combining Eqs. (\ref{MPD1})- (\ref{SCC}), we can obtain two conserved quantities associated with mass \(m\) and spin \(\mathcal{S}\) along the trajectory
\begin{eqnarray}
	P^{a}P_{a}=-m^{2}, \label{normalization of P}\\
	\mathcal{S}^2=\frac{1}{2}S^{ab}S_{ab}. \label{spin}
\end{eqnarray}
Following the convention adopted in \cite{Saijo1998}, we normalize the orbital affine parameter \(\tau\) as
\begin{eqnarray}
	P^{a}u_{a}=-m.\label{affine parameter}
\end{eqnarray}
With this choice, the four-vector \(u^{a}\) does not coincide with the normalized four-velocity and, in general, does not satisfy the normalization condition \(u^{a} u_{a} = -1\). To facilitate subsequent calculations, we introduce the spin four-vector \cite{Harms2016}
\begin{eqnarray} 
	S^{a}=-\frac{1}{2m}\epsilon^a_{\ bcd}P^{b}S^{cd}, \label{spin four-vector}
\end{eqnarray}
where \(\epsilon^a_{\ bcd}\) is the Levi-Civita tensor, and the spin four-vector is constrained by
\begin{equation}
	S^a S_a=\mathcal{S}^2. \label{constrain of spin four-vector}
\end{equation}
For simplicity, in this paper we restrict ourselves to trajectories lying in the equatorial plane (\(\theta=\pi/2\)). It follows immediately that \(P^{\theta} = 0\) and \(S^{b\theta} = 0\). Furthermore, substituting \(P^{\theta}\) and \(S^{b\theta}\) into Eq. (\ref{spin four-vector}), and using Eq. (\ref{constrain of spin four-vector}), we can identify
\begin{eqnarray}
	S_{t}=S_{r}=S_{\phi}=0,\,S_{\theta}=-\mathcal{S}\sqrt{g}.
\end{eqnarray}
To simplify the notation, we shorten \(g(r)\) to \(g\) (it should be noted that \(g\) does not represent the determinant of the metric). In subsequent sections, \(f\) and \(h\) will also denote the expressions given in Eqs. (\ref{f(r)}) and (\ref{h(r)}). Here \(\mathcal{S}\) not only characterizes the magnitude of the spin, but also encodes its orientation: for \(\mathcal{S}>0\) the particle spin is aligned with the orbital angular momentum, whereas for \(\mathcal{S}<0\) it is anti‑aligned. 
An equivalent form of the spin tensor (\ref{spin four-vector}) reads \cite{Harms2016}:
\begin{eqnarray}
	S^{ab}=-\frac{1}{m}\epsilon^{abcd}S_{c}P_d,
\end{eqnarray}
from which one can obtain the nonvanishing components
\begin{eqnarray}
	S^{tr}&=&-\frac{\mathcal{S} P_\phi}{m}\sqrt{\frac{h }{f g }},\\
	S^{t\phi}&=&\frac{\mathcal{S} P_r}{m}\sqrt{\frac{h }{f g }},\\
	S^{r\phi}&=&-\frac{\mathcal{S} P_t}{m}\sqrt{\frac{h }{f g }}.
\end{eqnarray}
A Killing vector \(\xi^a\) that describes spacetime symmetry can be associated with a conserved quantity \cite{Ehlers1977}
\begin{eqnarray}
	C = \xi^a P_a + \frac{1}{2}\,S^{ab} \nabla_a\xi_{b}. 
\end{eqnarray}
In a static, spherically symmetric spacetime, there are two Killing vectors: \((\partial/\partial t)^a\)
and \((\partial/\partial \phi)^a\), which are associated with the energy of the spinning particle and the total angular momentum, respectively,
\begin{eqnarray}
	\mathcal{E}&=&- P_t + \frac{\mathcal{S} P_\phi \,f' }{2 m}\sqrt{\frac{h }{f g }},\\
	\mathcal{J}&=&P_\phi-\frac{\mathcal{S} P_t \,g' }{2 m}\sqrt{\frac{h }{f g }}.
\end{eqnarray}
Using these conserved quantities, together with the normalization condition (\ref{normalization of P}),  the components of the four-momentum can be expressed as:
\begin{align}
	P_t&=\frac{ -4 m E f  g + 2 m s J  f'  \sqrt{f h g }}
	{4 f  g  - s^{2} h  f'  g' },\\
	P_\phi&=\frac{ 4 m J f  g - 2 m s E  g'  \sqrt{f h g }}
	{4 f  g  - s^{2} h  f'  g' },\\
	P_r^{2} &= -g_{rr}\left(g^{tt} P_t^{\,2}+g^{\phi\phi} P_\phi^{\,2} + m^{2}\right), \label{Pr}
\end{align}
where \(E=\mathcal{E}/m\), \(J=\mathcal{J}/m\) and \(s=\mathcal{S}/m\) denote the energy, angular momentum, and spin per unit mass of the particle, respectively. 

A point that merits attention is the reference value of the spin parameter. In realistic astrophysical scenarios, the spin of a Kerr black hole is bounded by $\mathcal{S} \leq m^2$, while that of a neutron star is typically even smaller. Furthermore, most celestial bodies would likely be disrupted by tidal forces as they approach stable orbits. Taking $\mathcal{S}= m^2$ as the theoretical upper limit for a spinning particle, the dimensionless spin is $s = \mathcal{S}/(mM) = m/M$. Since our model focuses on linear-order spin corrections for a particle moving in a fixed background without energy dissipation, the condition $m \ll M$ must be satisfied; otherwise, the comparable mass ratio would necessitate accounting for the particle's gravitational back-reaction on the spacetime. In this work, the mass of the central object has been normalized ($M = 1$), which theoretically implies $s \ll 1$. Nevertheless, in the subsequent numerical analysis, we adopt relatively large spin values unless otherwise specified. The reason for this choice is to more clearly demonstrate and accentuate the phenomenological differences induced by varying spin magnitudes, a methodology consistent with previous studies (e.g., \cite{Chen2024, Tan2024, Umarov2025}).

So far, we have successfully completed the detailed calculation of the four-momentum. The four-velocity most directly related to the orbit can be obtained from \cite{Saijo1998}
\begin{eqnarray}
	u^a =\frac{P^a}{m} + \frac{2S^{ab} R_{bcd e} P^c S^{de}}
	{m\left(4m^2 + R_{ijkl} S^{ij} S^{kl}\right)}. \label{velocity}
\end{eqnarray}
Its derivation is not particularly straightforward and is somewhat tedious, details can be found in \cite{Hackmann2014}.

%%%%%%%%%%%%%%%%%%%%%%%%%%%%%%%%%%%%%%  
\subsection{Timelike condition of the orbits}
%%%%%%%%%%%%%%%%%%%%%%%%%%%%%%%%%%%%%	
It follows from Eq. (\ref{MPD2}) that the directions of the four-momentum and four-velocity are generally inconsistent, which is essentially caused by the spin-gravity coupling effect. While Eq. (\ref{normalization of P}) ensures the timelike property of the four-momentum, the affine parameter selected in Eq. (\ref{affine parameter}) implies that the normalization condition of the four-velocity does not hold. Therefore, certain parameter ranges can cause the velocity of the particle to exceed the speed of light and its trajectory to become spacelike, which is physically unacceptable. To exclude such non-physical results, we introduce a superluminal constraint
\begin{equation}\label{timelikecondition}
	u^a u_a< 0,
\end{equation}
which is taken into account in all subsequent practical calculations.		

%%%%%%%%%%%%%%%%%%%%%%%%%%%%%%%%%%%%%  
\section{Effective potential and ISCO}\label{EFFECTIVE POTENTIAL AND ISCO}
%%%%%%%%%%%%%%%%%%%%%%%%%%%%%%%%%%%%%%%

%%%%%%%%%%%%%%%%%%%%%%%%%%%%%%%%%%%%%%
\subsection{Effective potential}
%%%%%%%%%%%%%%%%%%%%%%%%%%%%%%%%%%%%%%    
Since we focus on bound and (marginally) stable circular motion in the vicinity of a black hole, the spinning particles considered here neither escape to infinity nor plunge into the horizon. We restrict the motion to the equatorial plane. To identify the boundaries of its radial motion, the most straightforward approach would be to analyze the locations where \( u^r=0 \). However, a more convenient approach follows from the structure of Eq. (\ref{velocity}). A detailed evaluation shows that $$u^r = U(r)P^r.$$ Because the function $U(r)$ is nonvanishing outside the event horizon for the parameter ranges of interest, the turning points of the trajectory are equivalently determined by the condition $P^r=0$. Directly from Eq. (\ref{Pr}), we have:
\begin{align}
	\frac{(P^r)^2}{m^2}&=\frac{(\mathcal{A} E^2 + \mathcal{B} E + \mathcal{C})}{\mathcal{D}}  \nonumber\\
	&=\frac{\mathcal{A}}{\mathcal{D}}\left[E-\left(\frac{-\mathcal{B}+\sqrt{\mathcal{B}^2-4\mathcal{A}\mathcal{C}}}{2\mathcal{A}}\right)\right]\nonumber\\
	&\quad \times \left[E-\left(\frac{-\mathcal{B}-\sqrt{\mathcal{B}^2-4\mathcal{A}\mathcal{C}}}{2\mathcal{A}}\right)\right],
\end{align}
where
\begin{align}
	\mathcal{A} &= 4 f h  \left( 4g^2 - s^2 h g'^2\right), \\
	\mathcal{B} &= 16 s J h^{3/2}\sqrt{f g } \left( f g'  - g  f'  \right), \\
	\mathcal{C} &= h  \Big[ 4 J^2  g  \left( s^2 h f'^2 -4 f ^2\right) 
	- \left( s^2 h  f'  g'  -4 f  g \right)^2 \Big], \\
	\mathcal{D} &= \left( s^2 h f' g' -4 f g \right)^2.
\end{align}
Note that 
$\mathcal{D}$ is manifestly nonnegative, guaranteeing regular behavior of the radial equation away from singular points. The effective potential of relevance thus is defined as
\begin{eqnarray}
	V_{\text{eff}} &=& \frac{-\mathcal{B} + \sqrt{\mathcal{B}^2 - 4 \mathcal{A} \mathcal{C}}}{2 \mathcal{A}},
\end{eqnarray}
where the positive square root is chosen to correspond to a future-directed orientation. Bound motion requires the discriminant 
$\mathcal{B}^2 - 4 \mathcal{A} \mathcal{C} \geq 0$, ensuring that the effective potential remains real. It follows directly that when the particle’s energy \( E \) is equal to the effective potential, this marks the turning point of its radial motion. We introduce the orbital angular momentum parameter $l=J-s,$ which subtracts the intrinsic spin contribution $s$ from the conserved total angular momentum $J$.

\begin{figure}[!htbp]
	\centering
	\includegraphics[width=0.35\linewidth]{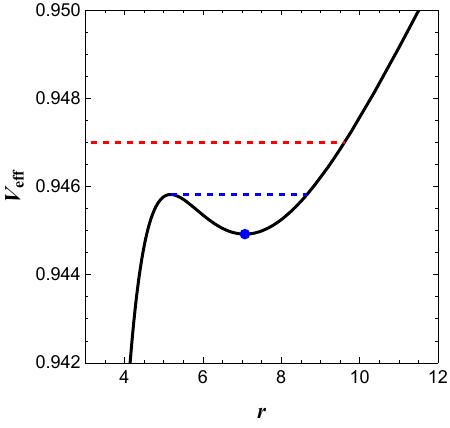}
	\caption{Effective potential \(V_{\text{eff}}\) as a function of \(r\) with \(b=2, \zeta=0.1, s=0.1, l=3.5\). The local maximum (blue dashed curve) defines the potential barrier separating bound and plunging trajectories, while the local minimum (blue dots) corresponds to stable circular orbits.}
	\label{fig:Veffs}
\end{figure}
As illustrated in Fig. \ref{fig:Veffs}, with appropriate parameter choices, we clearly observe that a potential barrier (corresponding to a local maximum) emerges in the effective potential near the horizon, and \( V_{\text{eff}} \to 1 \) as \( r \to +\infty \), this asymptotic limit follows from the Schwarzschild normalization at spatial infinity. If the particle's energy exceeds one, it will escape to infinity, corresponding to an unbound orbit. Similarly, if the local maximum (of the potential barrier) is less than one and the particle's energy surpasses this barrier, the particle will eventually fall into the black hole, which corresponds to the red dashed line in Fig. \ref{fig:Veffs}. Neither of these scenarios is desirable. For fixed parameters, the appropriate energy range for the particle is \( E_{\text{min}} < E < E_{\text{max}} \), where \( E_{\text{max}} \) denotes the local maximum at the potential barrier and \( E_{\text{min}} \) denotes the local minimum of \( V_{\text{eff}} \).

\begin{figure*}[t!]
	\centering
	\subfloat[$b=2$, $s=0.2$]{
		\includegraphics[width=0.32\textwidth]{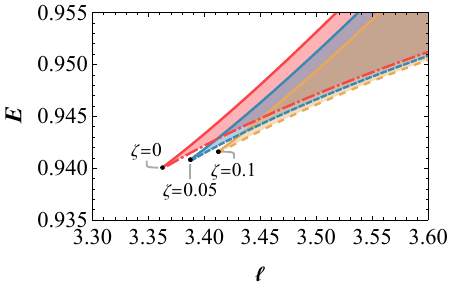}
	}
	\subfloat[$\zeta=0.1$, $s=0.2$]{
		\includegraphics[width=0.32\textwidth]{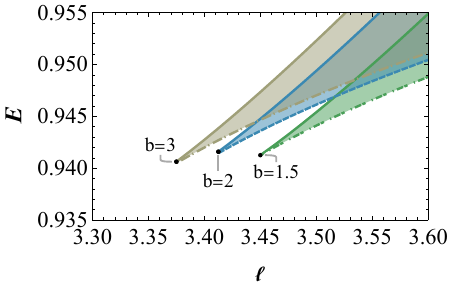}
	}
	\subfloat[$\zeta=0.1$, $b=2$]{
		\includegraphics[width=0.32\textwidth]{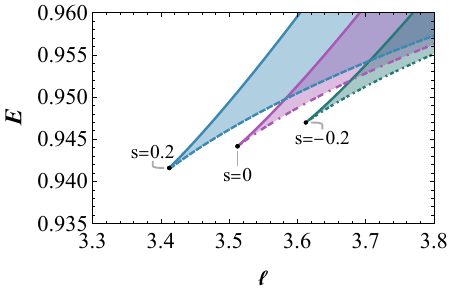}
	}
	\caption{  Allowed parameter region in the $(E,l)$ plane for bound motion. The shaded area corresponds to energies lying between the local maximum and minimum of the effective potential. The vertex of the shaded region marks the ISCO, where the two extrema merge. }
	\label{EJ}
\end{figure*}
\begin{figure*}[t!]
	\centering
	\subfloat[$b = 2$, $l = 3.45$, $s = 0.1$]{
		\includegraphics[width=0.3\textwidth]{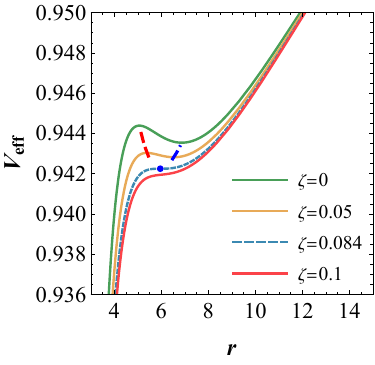}
		\label{fig:b2_Veff_r_Ag}
	}
	\subfloat[$\zeta = 0.1$, $s = 0.1$, $l = 3.5$]{
		\includegraphics[width=0.3\textwidth]{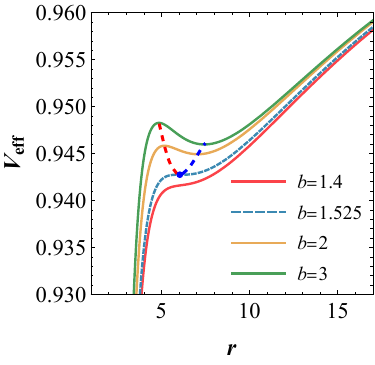}
		\label{fig:b2_Veff_r_b}
	}
	\subfloat[$\zeta = 0.1$, $b = 2$, $l = 3.5$]{
		\includegraphics[width=0.3\textwidth]{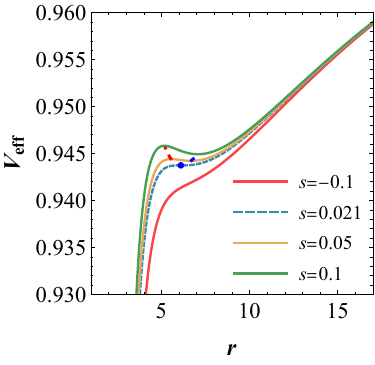}
		\label{fig:b2_Veff_r_s}
	}
	\caption{ The variation of $V_{\text{eff}}$ with respect to the radial coordinate \(r\). (a) Distinct $\zeta$ for fixed $b$, $s$ and $l$; (b) Distinct $b$ for fixed $\zeta$, $l$ and $s$;  (c) Distinct positive $s$ for fixed  $l$, $b$ and $\zeta$. The red dashed line traces the maxima (peaks) of the effective potential curves, while the dark blue dashed line marks the minima (valleys) across parameter variations.}
	\label{fig:b2_Veff_r}
\end{figure*}

Fig. \ref{EJ} provides an intuitive illustration in which the shaded region denotes the parameter range within which bound orbits can exist. For a fixed angular momentum, the upper energy bound decreases as \( \zeta \) increases. In contrast, increasing \( b \) causes the entire parameter range to shift to the left, with the upper bound of \( E \) instead increasing. Furthermore, the particle can carry higher energy when its spin is aligned with the orbital angular momentum.

To further clarify the influence of the model parameters, Fig. \ref{fig:b2_Veff_r} shows the variation of the effective potential with \(r\) for different parameter values. Here, as before, \( l = J - s \) denotes the orbital angular momentum. In Fig. \ref{fig:b2_Veff_r_Ag}, increasing \( \zeta \) causes the effective potential to decrease, whereas Fig. \ref{fig:b2_Veff_r_b} shows that the effective potential tends to rise with an increasing \(b\).  These trends indicate that stronger NLG correction terms result in a lower potential barrier. In Fig. \ref{fig:b2_Veff_r_s}, as the spin \( s \) varies from \(-0.1\) to \( 0.1 \), the effective potential also increases steadily, which demonstrates that spin can alter the stability of the particle’s orbit. In each panel, the blue dashed line corresponds to the cases where the maximum and minimum of the effective potential $V_{\text{eff}}$ coincide. Meanwhile, the dark blue solid points actually mark the ISCO, which will be introduced in the next subsection.
%%%%%%%%%%%%%%%%%%%%%%%%%%%%%%%%%%%%%%%%%%%%	
\subsection{ISCO}
%%%%%%%%%%%%%%%%%%%%%%%%%%%%%%%%%%%%%%%%%%%    
Here, we focus on the properties of ISCO. Its definition can be expressed via the conditions:
\begin{align}\label{ISCO}
	V_{\text{eff}}(r) = E, \quad \frac{dV_{\text{eff}}(r)}{dr} = 0, \quad \frac{d^2V_{\text{eff}}}{dr^2} = 0,
\end{align}
which correspond to a circular orbit at the inflection point of the effective potential. Despite its name as the ``innermost stable circular orbit", the mathematical conditions—with both the first and second derivatives of the effective potential vanishing—mean the orbit is actually unstable. When a particle on this orbit experiences small perturbations, it can hardly sustain circular motion and will eventually plunge into the black hole. However, we use this terminology to align with other papers.

\begin{figure*}[!htb]
	\centering
	\includegraphics[width=0.31\textwidth]{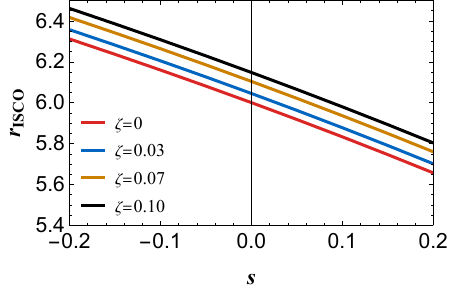}
	\includegraphics[width=0.333\textwidth]{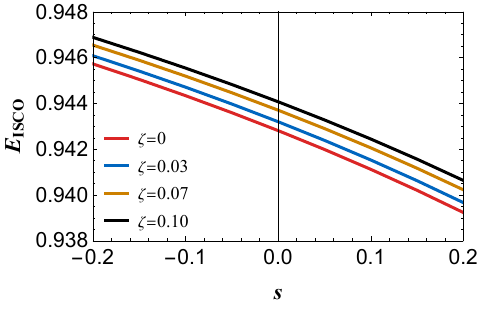}
	\includegraphics[width=0.325\textwidth]{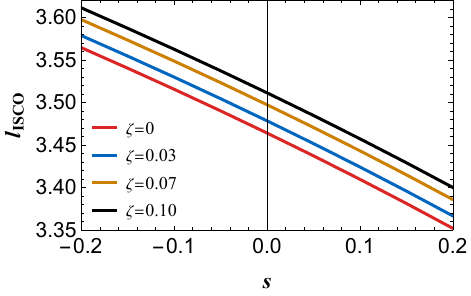}
	\caption{Dependence of the ISCO radius (left), energy (center), and orbital angular momentum (right) on the spin parameter $s$, for several values of the parameter $\zeta$, with $b=2$ fixed.}
	\label{fig:ISCO varyzeta}
\end{figure*}

These three Eqs. (\ref{ISCO}) allow us to solve for $\{r_{\text{ISCO}}, E_{\text{ISCO}}, l_{\text{ISCO}}\}$, provided the remaining nonlocal gravitational parameters are fixed. In Fig. \ref{fig:ISCO varyzeta}, we display the ISCO properties for different $\zeta$ values. For co-rotating particles ($sl > 0$), increasing spin $s$ reduces the ISCO radius, energy, and angular momentum. In contrast, for a fixed spin $s$, larger values of $\zeta$ lead to an increase in the ISCO radius, energy, and angular momentum for both co‑rotating and counter‑rotating orbits.
 
\begin{figure*}[!htb]
	\centering
	\includegraphics[width=0.31\textwidth]{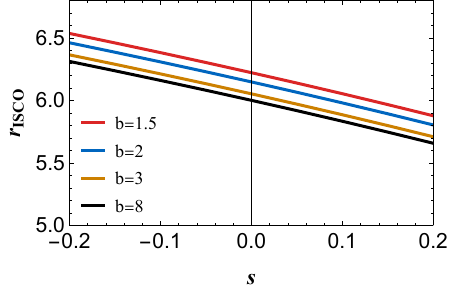}
	\includegraphics[width=0.333\textwidth]{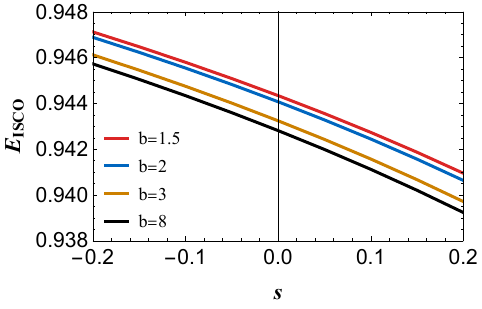}
	\includegraphics[width=0.325\textwidth]{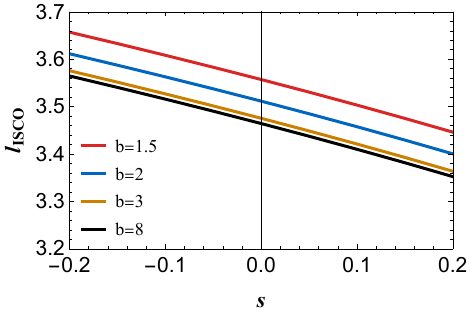}
	\caption{ISCO radius (left), energy (center), and orbital angular momentum (right) as functions of the spin parameter $s$ for different values of $b$, with $\zeta=0.1$ fixed. Increasing $b$ shifts the ISCO inward and lowers the corresponding energy, reflecting the influence of the nonlocal parameter on the near-horizon geometry.}
	\label{fig:ISCO varyb}
\end{figure*}

Fig. \ref{fig:ISCO varyb} illustrates the ISCO properties across different values of the parameter \( b \).  Increasing the spin \( s \) again decreases  the ISCO in radius, energy, and angular momentum, consistent with previous analysis. For a fixed \( s \), larger values of \( b \) lead to a smaller ISCO radius, energy and angular momentum. Additionally, the vertex of the shaded region in Fig. \ref{EJ} corresponds precisely to the ISCO values of energy and angular momentum, and its variation trends across different values of \(\zeta\), \(b\), and \(s\) align with the analysis presented here.


%%%%%%%%%%%%%%%%%%%%%%%%%%%%%%%%%%%%%%%%%%%%%%%    
\section{Periodic orbits and gravitational waves}\label{POandGW}
%%%%%%%%%%%%%%%%%%%%%%%%%%%%%%%%%%%%%%%%%%%%%%
Periodic orbits represent a special category of bound orbits, which are not only easier to handle but also serve as the reference base, such that any arbitrary orbit can be treated as a small perturbation of a precisely defined periodic orbit \cite{periodic_Levin_2008}. As such, the study of periodic orbits allows us to capture the key physical characteristics. Furthermore, discrepancies in spacetime geometry predicted by different gravitational theories are directly imprinted in the features of gravitational waveforms. This offers a powerful observational tool for testing and constraining various gravitational theories. In this section, we first present a brief introduction to periodic orbits and GWs, followed by a collective presentation of the numerical calculation results.

%%%%%%%%%%%%%%%%%%%%%%%%%%%%%%%%%%%%%%%%%%%%
\subsection{Periodic orbits}
%%%%%%%%%%%%%%%%%%%%%%%%%%%%%%%%%%%%%%%%%%%%%    
We adopt the classification scheme for periodic orbits introduced in \cite{periodic_Levin_2008} to examine whether NLG black holes can be distinctly differentiated from their Schwarzschild counterparts. The periodic motion of particles in the equatorial plane can be characterized by three integers, namely $(z, w, v)$:

$z$ (zoom): the number of "leaves" formed by each periodic orbit before closure;

$w$ (whirl): the number of additional circumferential revolutions executed by a particle near the perihelion during a single radial period (from one aphelion to the perihelion and then to the next aphelion);

$v$ (vertex): the number of vertices skipped by the particle when moving from the first aphelion to the next. 
\begin{figure*}[!tb]
	\centering
	\subfloat[$b = 2$,$s=0.1$,$E=0.97$]{
		\includegraphics[width=0.32\textwidth]{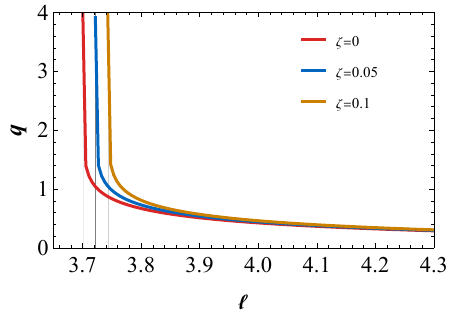} % 这里替换为你的文件路径
	}
	\subfloat[$\zeta = 0.1$,$s=0.1$,$E=0.97$]{
		\includegraphics[width=0.32\textwidth]{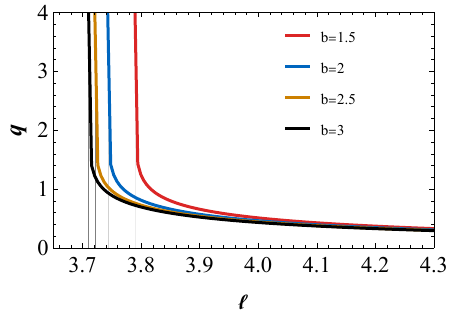} % 这里替换为你的文件路径
	}
	\subfloat[$b = 2$,$\zeta=0.1$,$E=0.97$]{
		\includegraphics[width=0.32\textwidth]{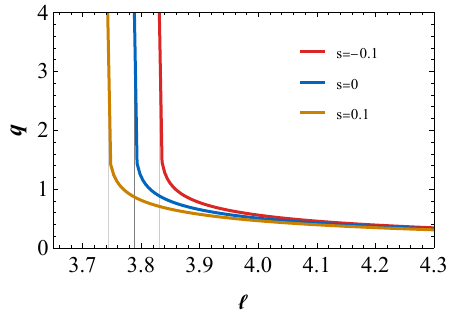} % 这里替换为你的文件路径
	}
	\quad
	\subfloat[$b = 2$,$s=0.1$,$l=3.8$]{
		\includegraphics[width=0.32\textwidth]{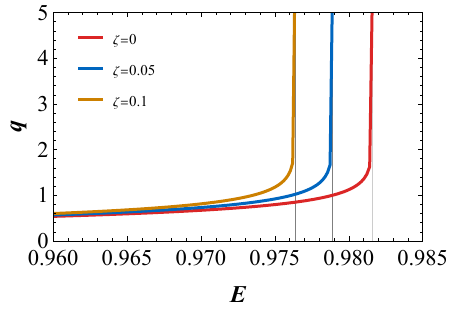} % 这里替换为你的文件路径
	}
	\subfloat[$\zeta = 0.1$,$s=0.1$,$l=3.8$]{
		\includegraphics[width=0.32\textwidth]{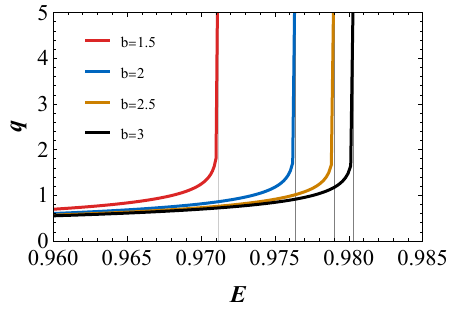} % 这里替换为你的文件路径
	}
	\subfloat[$b=2, \zeta = 0.1$, $l=3.8$]{
		\includegraphics[width=0.32\textwidth]{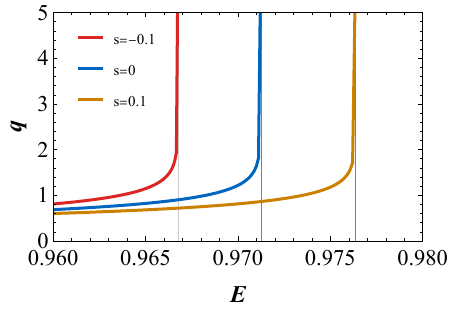} % 这里替换为你的文件路径
	}
	\caption{{\bf{Upper panels:}} Dependence of the parameter $q$ and the orbital angular momentum $l$ for different values of: (a) \(\zeta\); (b) \(b\); (c) \(s\). {\bf{Lower panels:}} Dependence of the parameter $q$ and the energy $E$ under the same parameter variations: (d) varying \(\zeta\); (e) varying \(b\); (f) varying \(s\).}
	\label{fig:PO}
\end{figure*}
For orbits with $z\ge2$, there exist multiple aphelia that make up the vertices of a polygon. These vertices are numbered sequentially in the direction of orbital rotation. For example, the initial aphelion is denoted as $v=0$; $v=1$ means that the particle travels directly from the initial aphelion to the first vertex, while $v=2$ implies that the particle moves from the initial aphelion to the second vertex, and so forth.

Notably, the zoom-whirl behavior is a physically realizable orbital feature that modulates gravitational waveforms even with gravitational radiation dissipation \cite{Zoom-Whirl_Healy_2009}. The integers $(z, w, v)$ can be related to a rational number $q$, which satisfies the following relation for periodic orbits \cite{periodic_Levin_2008}
\begin{align} \label{qFun}
	q = \frac{\Delta \phi}{2\pi} - 1 = w + \frac{v}{z}.
\end{align}
Here, $\Delta \phi$ denotes the azimuthal angle traversed by the particle over a complete radial period. It can be derived by integrating the relation between azimuthal and radial motion, which reads:
\begin{align}\label{Deltaphi}
	\Delta \phi = 2\int_{r_1}^{r_2} \left(\frac{d\phi}{dr}\right) dr,
\end{align}
where \(r_1\) and \(r_2\) denote the perihelion and aphelion, respectively. Using Eq. (\ref{velocity}), we have 

\begin{align}\label{integrand}
		\frac{d\phi}{dr} = \frac{d\phi}{d\tau} \frac{d\tau}{dr} = \frac{u^\phi}{u^r},
\end{align}
which depends on the  energy $E$, orbital angular momentum $l$, spin $s$, as well as $\zeta$ and $b$. By combining Eqs. (\ref{qFun})-(\ref{integrand}), once the values of $s$, $\zeta$ and $b$ are specified, the correlation relationships between $q$ and $E$, $l$ can be derived separately.

Detailed calculations are presented in Fig. \ref{fig:PO}. In the upper panels, with $E$ kept constant, $q$ increases gradually and eventually diverges as $l$ decreases. In the lower panels, with $l$ fixed, $q$ rises slowly and diverges as $E$ approaches its maximum value. Moreover, for a fixed value of \(q\), an increase in \(\zeta\) leads to a larger orbital angular momentum but a smaller energy. In contrast, a larger \(b\) results in a smaller orbital angular momentum but a larger energy, showing a competitive trend against the effect of \(\zeta\). When the spin is aligned with the orbital angular momentum, the minimum value of \( l \) decreases while the maximum value of \( E \) increases. This behaviour indirectly implies that co-rotating particles are more likely to form periodic orbits—meaning that they tend to stabilize the orbit. Conversely, particles with spin anti-aligned with the orbital angular momentum reduce orbital stability. These trends are consistent with those discussed in Figs. \ref{EJ} and \ref{fig:b2_Veff_r}, and such characteristics can serve as a basis for distinguishing between different NLG black holes.

\begin{figure}[!tbp]
	\centering
	\begin{subfigure}[b]{\linewidth}
		\centering
		\includegraphics[width=0.32\textwidth]{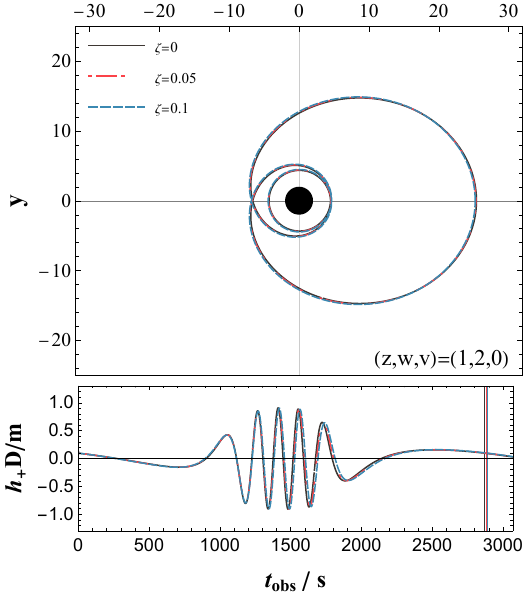}
		\includegraphics[width=0.32\textwidth]{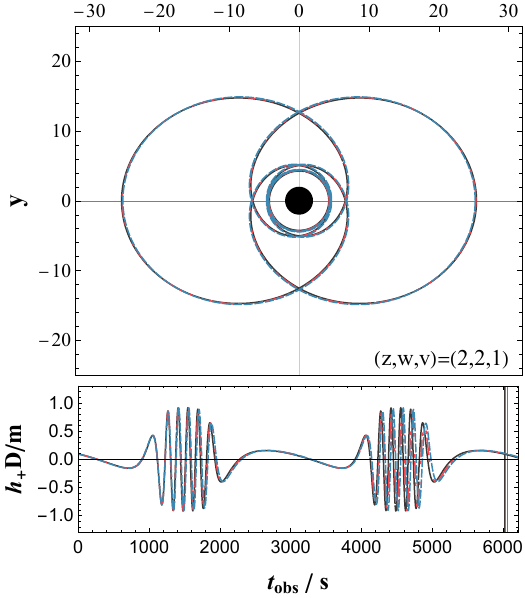}
		\includegraphics[width=0.32\textwidth]{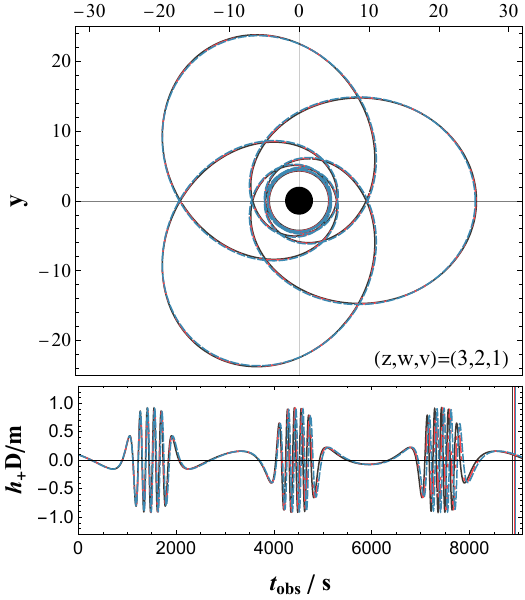}
		\caption{$b=2$, $s=0.1$, $E=0.97$}
		\label{fig:group1}
	\end{subfigure}
	
	\begin{subfigure}[b]{\linewidth}
		\centering
		\includegraphics[width=0.32\textwidth]{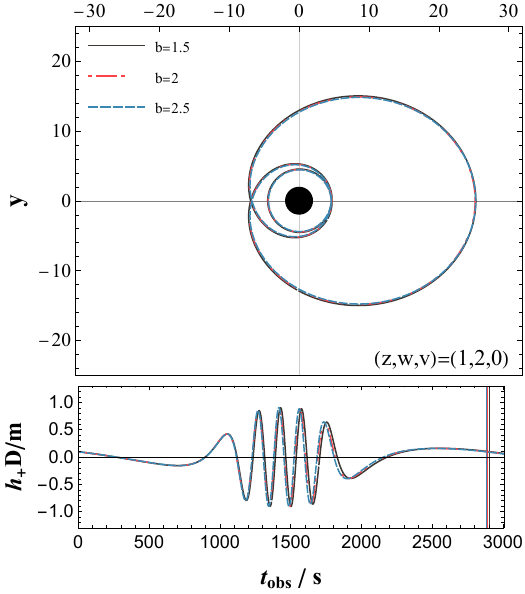}
		\includegraphics[width=0.32\textwidth]{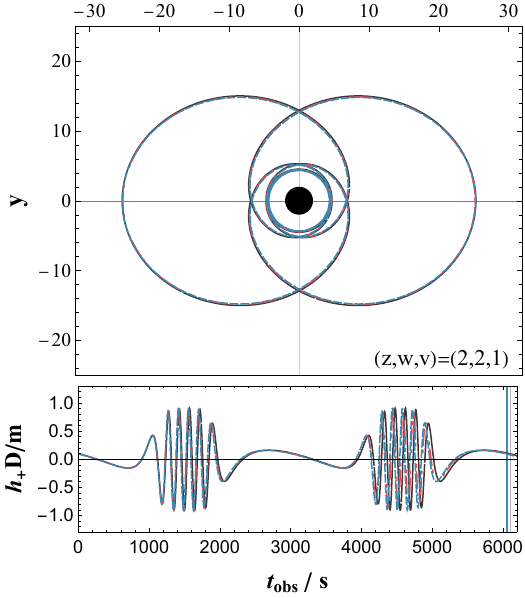}
		\includegraphics[width=0.32\textwidth]{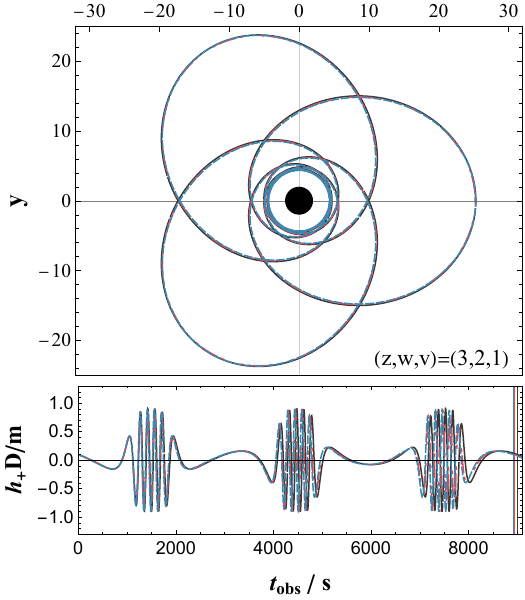}
		\caption{$\zeta=0.1$, $s=0.1$, $E=0.97$}
		\label{fig:group2}
	\end{subfigure}
	
	\begin{subfigure}[b]{\linewidth}
		\centering
		\includegraphics[width=0.32\textwidth]{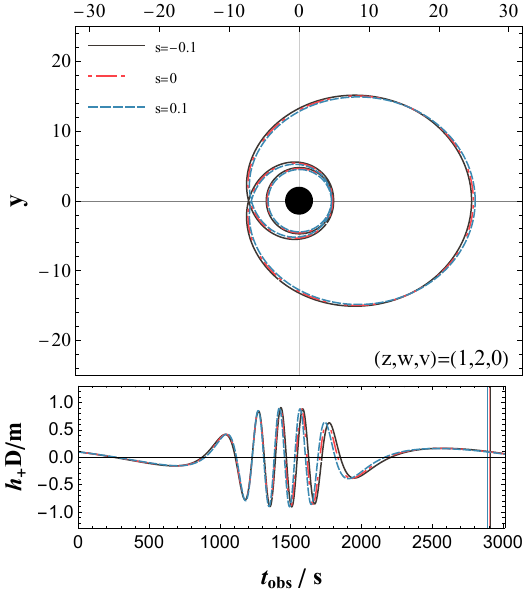}
		\includegraphics[width=0.32\textwidth]{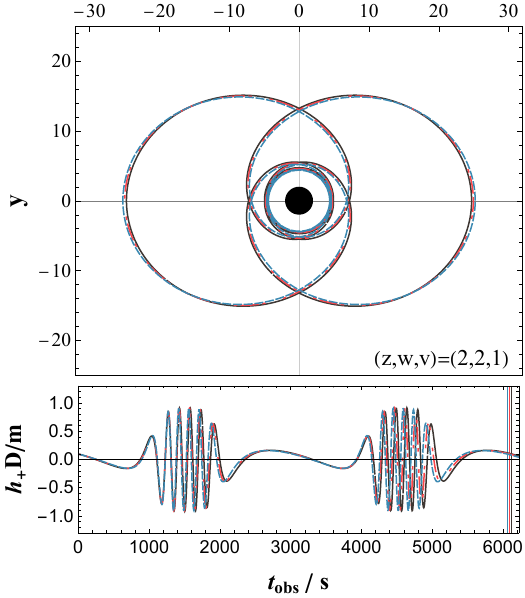}
		\includegraphics[width=0.32\textwidth]{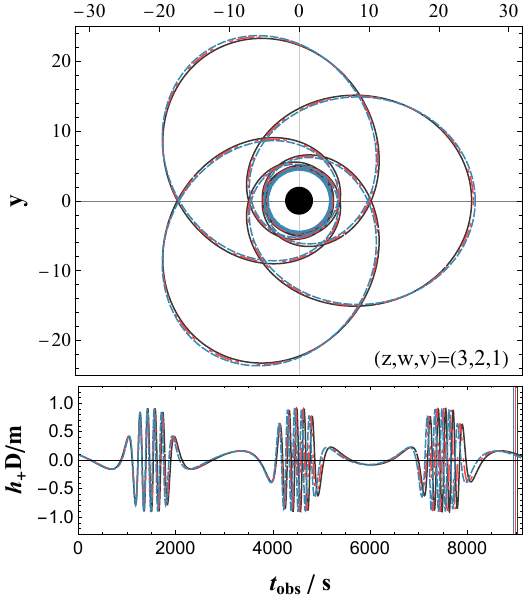}
		\caption{$\zeta=0.1$, $b=2$, $E=0.97$}
		\label{fig:group3}
	\end{subfigure}
	\caption{Periodic trajectories and the corresponding gravitational waveforms for different values of: (a) \(\zeta\), (b) \(b\), and (c) spin \(s\). The central black hole mass is set to \(M = 10^6 M_\odot\), the compact object mass to \(m = 10M_{\odot}\), and the luminosity distance to \(D = 1\) Gpc. Additionally, we adopt \(\Theta = \pi/2\) and \(\Phi = \pi/3\), which results in \(h_{\times} = 0\). The vertical line in the waveform marks the time corresponding to one complete orbital period.}
	\label{fig:OBandGW}
\end{figure}

The calculation of specific orbits depends on Eq. (\ref{velocity}). This relationship depends explicitly on \( r \) but implicitly on \( \tau \) (i.e., \( u^a(r(\tau)) \)). The system of differential equations for the trajectory is given as follows
\begin{equation}
		\begin{aligned}
			\frac{dt}{d\tau}&=u^t, \\
			\frac{d^2r}{d\tau^2}&=\frac{du^r}{dr} \cdot u^r,\\
			\frac{d\theta}{d\tau}&=0 ,\\
			\frac{d\phi}{d\tau}&=u^\phi.
		\end{aligned}
\end{equation}
Given the initial conditions, the orbit parameterized by the affine parameter \(\tau\) can be derived. However, for the convenience of subsequent GW calculations, we aim here to use the particle's coordinate time  \(t\)  instead of  \(\tau\)  as the parameter along its trajectory. The transformed system reads:
\begin{equation}
		\begin{aligned}\label{orbital}
			r''(t) &= \frac{d^2 r}{d\tau^2} \cdot \left( \frac{d\tau}{dt} \right)^2 + \frac{dr}{d\tau} \cdot \frac{d^2 \tau}{dt^2} \\
			&= \frac{u^r}{\left(u^t\right)^2} \cdot \frac{du^r}{dr}- \frac{(r'(t))^2}{u^t}\cdot \frac{du^t}{dr}, \\
			\theta'(t)&=0 ,\\
			\phi'(t)&= \frac{d\phi}{d\tau} \cdot \frac{d\tau}{dt} = \frac{u^\phi}{ u^t},
		\end{aligned}
\end{equation}
where the prime denotes differentiation with respect to $t$. The initial conditions are chosen as
\begin{align}\label{initial condition}
	r'(0) = 0, \quad r(0) = r_2, \quad \theta(0)=\pi/2, \quad \phi(0) = 0,
\end{align}
The functions \(r(t)\) and \(\phi(t)\) are obtained by numerically solving the above equations. Projecting these solutions onto the two-dimensional plane via $x=r(t)\cos\phi(t)$ and $y=r(t)\sin\phi(t)$ yields the trajectories of periodic orbits. We present our computational results together with the gravitational waveforms in Fig. \ref{fig:OBandGW} and discuss them in detail in Section \ref{NS}.

%%%%%%%%%%%%%%%%%%%%%%%%%%%%%%%%%%%%
\subsection{Gravitational waveforms from periodic orbits}
%%%%%%%%%%%%%%%%%%%%%%%%%%%%%%%%%%%%% 

In realistic astrophysical scenarios, EMRIs consist of a compact object of much smaller mass orbiting a supermassive black hole. Here, we model the smaller object as a spinning test particle and compute waveforms from conservative periodic orbits using the numerical kludge approach described in \cite{Babak2007}. While this is a phenomenological approach, empirically, as long as the orbital perihelion is not too close to the black hole, the numerically kludge waveforms rapidly generated by this method achieve extremely high overlap with the accurate Teukolsky-based waveforms \cite{Glampedakis2005}. Radiation reaction is not included, and the orbital parameters are held fixed throughout the calculation. In particular, the one-year waveforms used below are constructed by periodically extending these fixed-parameter trajectories, rather than by evolving an inspiral. The resulting mismatch analysis is therefore intended as an order-of-magnitude sensitivity estimate.

First, we place the black hole at the coordinate origin and project the orbits in Boyer-Lindquist coordinates onto the Cartesian coordinates of a pseudo-flat spacetime
\begin{align}
	x_p = r\sin\theta\cos\phi,\quad y_p = r\sin\theta\sin\phi,\quad z_p = r\cos\theta.
\end{align}
Here, the subscript \( p \) denotes the Cartesian coordinates of the particle (i.e., \( (t_p, \vec{x}_p) \)), and \(t_p\)  is in fact identified with the orbital time parameter  \(t\) in Eqs. (\ref{orbital}). This formulation facilitates the application of wave-generation formulas derived for flat spacetime. In the weak-field approximation, the metric can be decomposed as
\begin{align}
	g_{ab}=\eta_{ab}+h_{ab},
\end{align} with \(\eta_{ab}\) denoting the Minkowski metric and \(h_{ab}\) the metric perturbation. We may then define the trace-reversed metric perturbation as
\begin{align}
	\bar{h}^{ab}=h^{ab}-\frac{1}{2}\eta^{ab}h,
\end{align}
where \(h=\eta^{cd}h_{cd}\) is the trace of \(h_{ab}\). Upon imposing the Lorentz gauge condition
\(\partial_{a}\bar{h}^{ab}=0\), we arrive at the linearized field equations:
\begin{align}
	\partial^{c}\partial_{c}\bar{h}^{ab}=-16\pi T^{ab}.
\end{align}
For a point particle moving along a trajectory, the energy-momentum tensor in flat spacetime can be written as
\begin{align}
	T^{ab}=m\cdot \frac{dx_{p}^a}{dt_p}\frac{dx_{p}^b}{dt_p}\delta^3(\vec{x}-\vec{x}_p),
\end{align}
In the slow-motion limit, GWs arise exclusively from the quadrupole moment contribution:
\begin{align}
	I^{ab}(t_p)&=\int x^a x^b T^{00}d^3x
	=m x_p^a x_p^b,
\end{align}
and the resulting waveform reads
\begin{align} \label{QFGW}
	\bar{h}^{ab}(t_{obs},\vec{x}_{obs})&= \frac{2}{D} \frac{d^2 I^{ab}(t_p)}{dt_{obs}^2},
\end{align}    
with \(D\) the luminosity distance from the source to the observer. We denote the observer’s position by (\(t_{obs},\vec{x}_{obs}\)), and \(t_{\text{obs}} = t_p + |\vec{x}_{\text{obs}} - \vec{x}_p|\) where the second term on the right-hand side accounts for the light-travel time between the particle and the observer. It is treated as a constant during the particle's motion. We now consider an observer located at the source's latitude and azimuthal angle (\(\Theta, \Phi\)), and project \(\bar{h}^{ab}\) onto the spherical coordinates, with its non-vanishing components given by:
\begin{equation}
	\begin{aligned}
		h^{\Theta\Theta} &= \cos^2\Theta\bigl( \bar{h}^{xx}\cos^2\Phi + \bar{h}^{xy}\sin2\Phi + \bar{h}^{yy}\sin^2\Phi \bigr),\\ %+ \bar{h}^{zz}\sin^2\Theta - \sin2\Theta\bigl( \bar{h}^{xz}\cos\Phi  + \bar{h}^{yz}\sin\Phi \bigr), \\
		h^{\Phi\Theta} &= \cos\Theta\biggl( -\frac{1}{2} \bar{h}^{xx}\sin2\Phi + \bar{h}^{xy}\cos2\Phi + \frac{1}{2} \bar{h}^{yy}\sin2\Phi \biggr), \\ %+ \sin\Theta\bigl( \bar{h}^{xz}\sin\Phi - \bar{h}^{yz}\cos\Phi \bigr), \\
		h^{\Phi\Phi} &= \bar{h}^{xx}\sin^2\Phi - \bar{h}^{xy}\sin2\Phi + \bar{h}^{yy}\cos^2\Phi.
	\end{aligned}
\end{equation}

This derivation has taken into account the fact that \(\bar{h}^{az}=0\). In the standard transverse-traceless (TT) gauge, the observable gravitational waveform is written as
\begin{align}
	h_{TT}^{ab}=\frac{1}{2}\begin{pmatrix}
		0 & 0 & 0 \\
		0 & h_+ & h_\times \\
		0 & h_\times & -h_+
	\end{pmatrix}
\end{align}
where \(h_+\) and \(h_\times\) are the two independent polarization states of GWs, with their explicit expressions given by
\begin{equation}
	\begin{aligned}
		h_+ &= h^{\Theta\Theta} - h^{\Phi\Phi}, \\
		h_\times &= 2 h^{\Phi\Theta}.
	\end{aligned}
\end{equation}
This set of operations achieves the matching of orbital parameters to gravitational waveforms.

To quantify the distinguishability between two gravitational waveforms, we introduce the noise-weighted inner product~\cite{Flanagan1998},
	\begin{align}
		\langle h_1 \mid h_2 \rangle
		=
		4\,\mathrm{Re}
		\int_{f_{\min}}^{f_{\max}}
		\frac{\tilde{h}_1(f)\tilde{h}_2^{*}(f)}
		{S_n(f)}\,df ,
	\end{align}
	where $\tilde{h}_1(f)$ and $\tilde{h}_2(f)$ are the frequency-domain waveforms obtained using the discrete Fourier transform.
	A single closed periodic orbit was first computed at high precision and then periodically extended over the one-year observation interval. The time-domain waveforms are sampled uniformly over one year using $2^{22}=4\,194\,304$ points. This corresponds to a sampling interval of $\Delta t=7.51877\,{\rm s}$, a sampling frequency of $0.13300\,{\rm Hz}$. A Tukey window with parameter
	$\alpha=0.1$ is applied before the Fourier transform. The inner product is evaluated over
		\(f_{\min}=10^{-4}\,{\rm Hz}\) and \(f_{\max}=5\times10^{-2}\,{\rm Hz}\).
	Here $S_n(f)$ denotes the one-sided LISA noise power spectral density, including both the instrumental noise and the one-year Galactic-binary confusion foreground \cite{Robson:2018ifk}.
	
	To remove these unphysical differences, we define the maximized faithfulness by maximizing the overlap over a constant time shift \(t_c\) and phase offset \(\psi_c\) \cite{Zhang:2026ujb}:
	\begin{align}
		F_{\max}(h_1,h_2)
		=
		\max_{t_c,\psi_c}
		\frac{
			\langle h_1\mid h_2(t_c,\psi_c)\rangle
		}{
			\sqrt{
				\langle h_1\mid h_1\rangle
				\langle h_2\mid h_2\rangle
			}
		},
	\end{align}
	where the corresponding frequency-domain transformation is
	\begin{align}
		\tilde{h}_2(f;t_c,\psi_c)
		=
		\tilde{h}_2(f)
		\exp\!\left[i(2\pi f t_c+\psi_c)\right].
	\end{align}
	The mismatch is then defined as
	\begin{align}
		\mathcal{M}(h_1,h_2)=1-F_{\max}(h_1,h_2).
	\end{align}
	We have verified that the mismatch remains numerically stable under refinements of both the orbital-integration accuracy and the waveform sampling resolution. As a reference criterion for the sensitivity estimate, we adopt
	\begin{align}
		\mathcal{M}_c=\frac{N}{2\rho^2},
	\end{align}
	where $\rho$ is the signal-to-noise ratio and $N$ is the number of independent intrinsic waveform parameters. In the present estimate, we take $N=6$, corresponding to $M$, $m$, $E$, $s$, $\zeta$, and $b$. The angular momentum $\ell$ is not counted as an independent parameter, because it is determined by these parameters together with the specified periodic-orbit class $(z,w,v)$. For the representative value $\rho=20$, the resulting threshold is $\mathcal{M}_c=0.0075$. This criterion is used only as a reference for an order-of-magnitude sensitivity estimate within the fixed-orbit waveform model. It does not establish detectability for an evolving EMRI, since radiation reaction and a complete parameter-degeneracy analysis are not included.

%%%%%%%%%%%%%%%%%%%%%%%%%%%%%%%%%%%%%%%%%%%%%%%%   
\subsection{Numerical results}\label{NS}
%%%%%%%%%%%%%%%%%%%%%%%%%%%%%%%%%%%%%%%%%%%%%%%%

\begin{figure*}[!tbp]
	\centering
	
	\begin{subfigure}[c]{0.53\linewidth}
		\centering
		\includegraphics[width=1.11\linewidth]{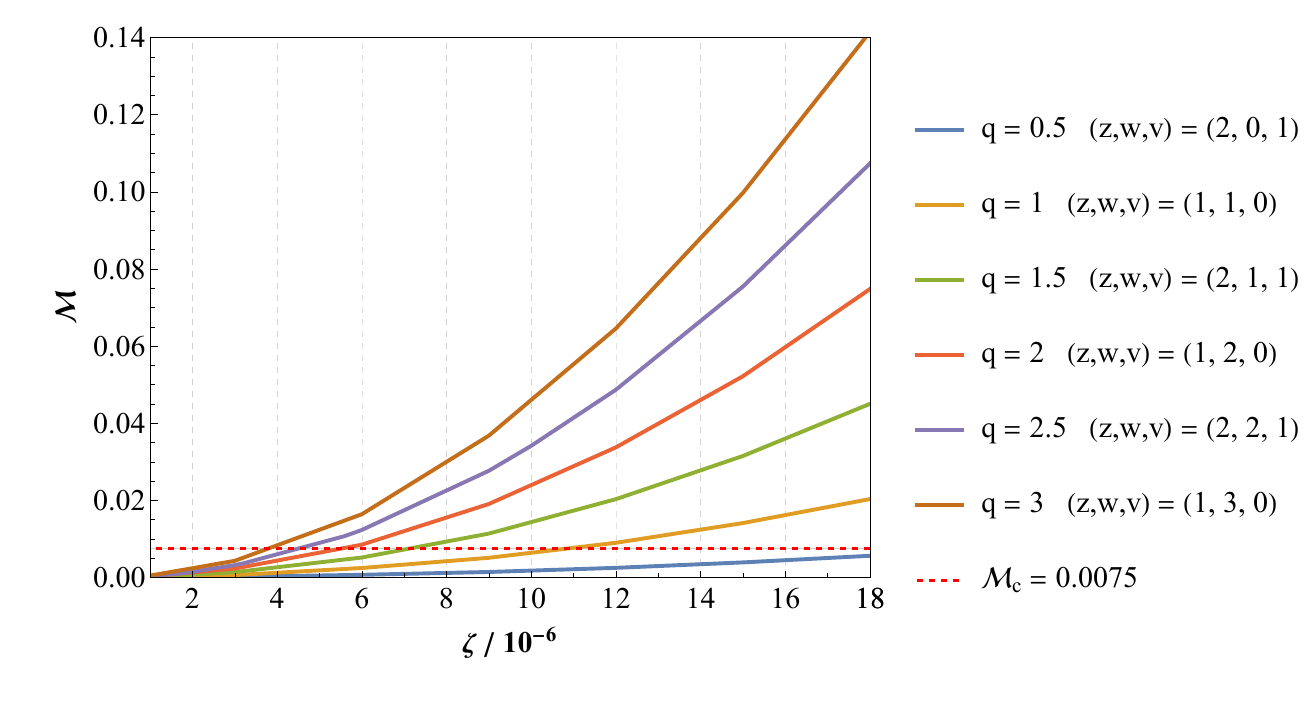} 
		\caption{Mismatch $\mathcal{M}$ versus $\zeta$ for different periodic orbits $(z,w,v)$ with \(b=2\) .}
		\label{fig:mismatchvszeta}
	\end{subfigure}
	\hfill 
	\begin{subfigure}[c]{0.4\linewidth}
		\centering
		\includegraphics[width=1\linewidth]{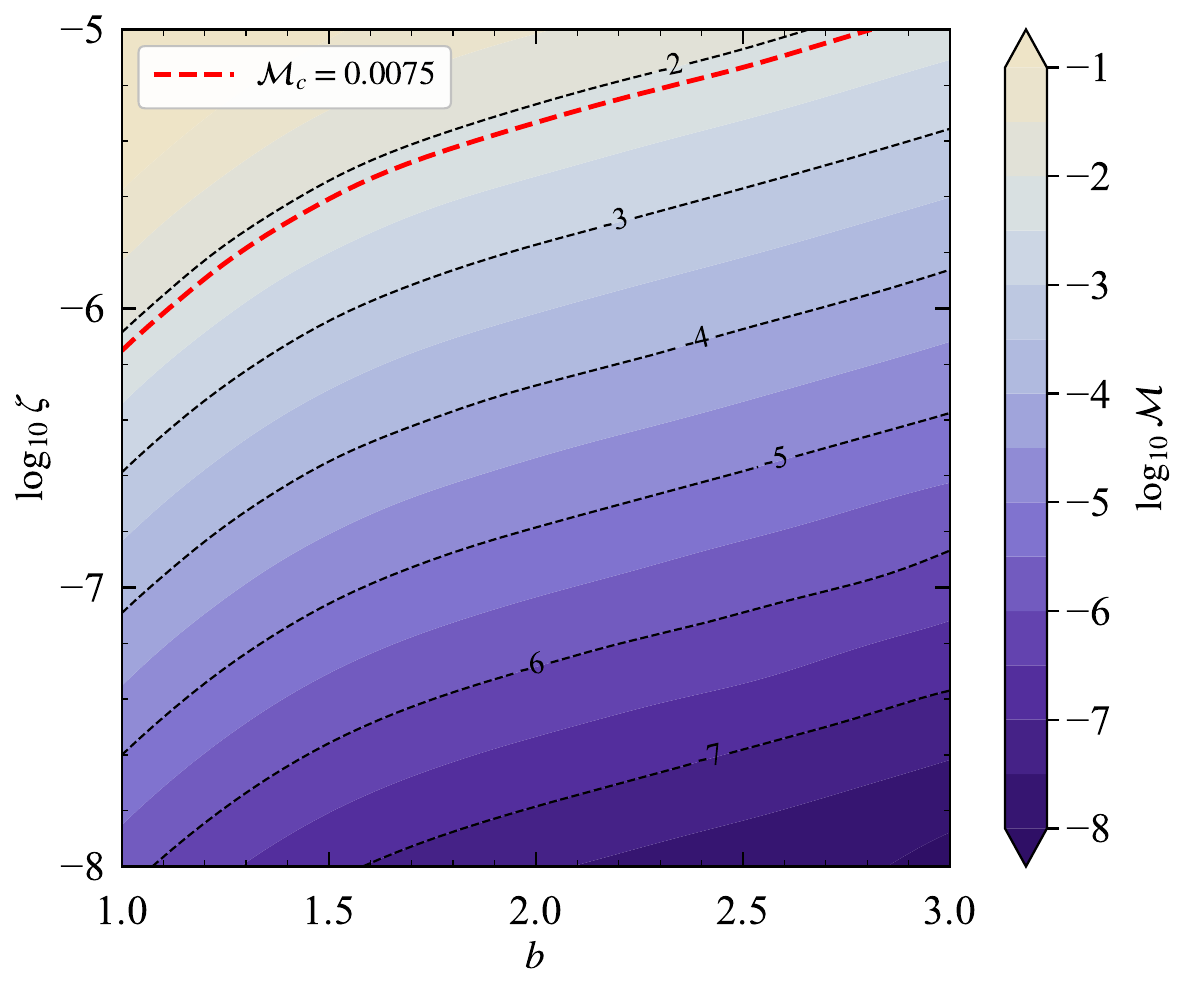}
		\caption{Contour plot of $\log_{10} \mathcal{M}$ in the $(b, \zeta)$ plane for the periodic orbit $(z, w, v) = (2, 2, 1)$.}
		\label{fig:mismatch contour}
	\end{subfigure}
	
	\caption{Mismatch in $h_{+}$ between NLG and Schwarzschild waveforms constructed by periodically extending conservative orbits over a one-year interval. We consider an EMRI system with the following parameters: \(M = 10^6 M_\odot\), \(m = 10M_{\odot}\), \(s=10^{-5}\), \(E=0.97\), \(D = 1\) Gpc. Additionally, we adopt \(\Theta=\pi/2,\Phi=\pi/3\). The red line corresponds to the mismatch \(\mathcal{M} = 0.0075\)}
	\label{fig:gw_analysis}
\end{figure*}   

In Fig. \ref{fig:OBandGW}, we present three representative sets of periodic orbits together with their corresponding gravitational waveforms, illustrating the effect of the parameters: \(\zeta\), \(b\), and the spin \(s\). We note that variations in these parameters modify the orbital geometry. These changes, which appear primarily as shifts in the positions of perihelion and aphelion, are directly imprinted onto the emitted gravitational waveform.  

In particular, relative to the Schwarzschild case an increase in \(\zeta\) leads to a distinct phase delay in the waveform. The parameter \(b\) exerts a relatively weak influence on both the orbit and the waveform, yet our quantitative analysis reveals that increasing \(b\) results in a slight phase advance. Regarding the particle spin, when it is co-rotating, the waveform exhibits a phase advance, whereas a counter-rotating particle produces a phase delay—reflecting the modulation of orbital dynamics by spin-curvature coupling.  

The parameter differences in the waveform examples are chosen to illustrate their effects on orbital geometry and waveform phase. Smaller differences in orbital frequencies can accumulate phase offsets over many cycles, thereby increasing the mismatch between the periodically extended waveforms.
	
Fig. \ref{fig:gw_analysis} presents the mismatch between Schwarzschild and NLG waveforms constructed over a one-year interval with fixed orbital parameters. Since radiation reaction is neglected, we interpret the results as an order-of-magnitude sensitivity estimate rather than a definitive detectability prediction. We use $\mathcal{M}_c=0.0075$, introduced above, as a reference criterion. Fig. \ref{fig:mismatchvszeta} shows that, for fixed $b=2$, larger values of $q$ correspond to smaller values of $\zeta$ required to reach the reference criterion. This indicates greater waveform sensitivity to NLG effects for orbits with higher $q$, consistent with the trend reported in \cite{Gong2025}. Fig. \ref{fig:mismatch contour} displays the mismatch in the $(b,\zeta)$ parameter plane. The region above the red curve corresponds to $\mathcal{M}>\mathcal{M}_c$. For a fixed $\zeta$, increasing $b$ reduces the mismatch, indicating weaker waveform sensitivity to the nonlocal correction.

%%%%%%%%%%%%%%%%%%%%%%%%%%%%%%%%%%%%%%%%%%%%%%%%%%%%%%%%%%%%%%%%%%%%%%%%%%%%%%%%%%%%%%%%%%%%%%%%%%%%%%%%%%%%%%%%%%%%%%%%%%%%%%%%%%%%%%%%%%%%%%%%%%%%%%%%%%%%%%%%%%%%%%%%%%%%%%%%%%%%%%%

\section{Conclusion and discussion}\label{conclusion}
%%%%%%%%%%%%%%%%%%%%%%%%%%%%%%%%%%%%%%%%%%%%%%%%%%%%%%%%%%%%%%%%%%%%%%%%%%%%%%%%%%%%%%%%%%%%%%%%%%%%%%%%%%%%%%%%%%%%%%%

This work presents a systematic analysis of the orbital dynamics of spinning particles and the corresponding GW signals in the static, spherically symmetric black hole solution derived from a modified DW NLG framework. The aim is to uncover the distinctive imprints of nonlocal gravitational corrections on spacetime structure. First, by employing the MPD equations together with the supplementary Tulczyjew spin condition, we derived the equations of motion for spinning particles in this NLG background. A timelike constraint was further imposed to eliminate unphysical superluminal trajectories.

We constructed an effective potential for radial motion and analyzed the permissible ranges of energy and angular momentum that support bound orbits. We computed key properties of the ISCO. Our results show that a larger nonlocal perturbation parameter \(\zeta\) lowers the potential barrier and increases the ISCO radius, whereas a larger value of \(b\) raises the barrier and decreases the ISCO radius. Particle spin also modulates orbital stability: for spin aligned with the orbital angular momentum, the potential barrier is raised and the ISCO radius reduced; the opposite behavior is observed for anti-aligned spin, which manifests a distinct spin–orbit coupling effect.

Adopting the \((z,w,v)\) classification scheme, we characterized the various kinds of periodic orbits and computed their associated gravitational waveforms. We found that nonlocal corrections induce phase shifts in the waveforms: increasing \(\zeta\) gives rise to phase delays, whereas increasing \(b\) causes phase advance. Similarly, co-rotating particles also produce phase advance. Within the fixed-orbit waveform model extended over one year, the mismatch results suggest an order-of-magnitude sensitivity to $\zeta\sim10^{-6}$ for $b=2$ in the configurations considered, while larger values of $b$ reduce the mismatch. These results are sensitivity estimates rather than definitive detectability predictions.

Radiation reaction is not included in the present calculation, and the orbital parameters are held fixed over the one-year interval. We have therefore not quantified the effects of the gravitational-wave-induced losses of energy and angular momentum on the secular orbital evolution and, consequently, on the accumulated waveform phase and mismatch. A definitive assessment of detectability requires the self-consistent inclusion of radiation reaction, together with dedicated statistical studies such as Fisher-matrix forecasting and Bayesian analysis \cite{Barack2003}. Such an analysis is considered as a future work. 

Future research directions can be extended to the study of black hole chaotic behavior. For instance, spinning particles may exhibit chaotic motion in black hole spacetimes \cite{Suzuki1997, Hartl2003, Zelenka2020}. Given that chaos is highly sensitive to the underlying spacetime geometry, diagnostic tools such as Lyapunov exponents and Poincaré sections can amplify subtle differences between NLG black holes and GR black holes, thereby offering a promising avenue for discriminating among different NLG theories.

\acknowledgments

This work is supported by the National Natural Science Foundation of China (NSFC) with Grant No. 12275087. MB is supported by Proyecto Interno  UCM-IN-25202 l\'inea regular and  FONDECYT grant 1262452.

\bibliographystyle{jhep}
\bibliography{ref}

\end{document}